\documentclass[aps, prd, showpacs, superscriptaddress, ctexart, nofootinbib, twocolumn]{revtex4}

\usepackage{amssymb, amsmath, bm, dcolumn, epsf, graphicx, latexsym, slashed, simplewick}
\pdfoutput=1
\usepackage{eurosym}
\usepackage{verbatim}
\usepackage{color}

\usepackage{hyperref}
\usepackage[all]{hypcap}

\def\be{\begin{equation}}
\def\ee{\end{equation}}
\def\bea{\begin{eqnarray}}
\def\eea{\end{eqnarray}}
\usepackage{graphicx}
\usepackage{dcolumn}
\usepackage{bm}
\begin{document}

\title{Note on Structure Formation from Cosmic String Wakes}

\author{Francis Duplessis}\thanks{francis.duplessis@mail.mcgill.ca}
\author{Robert Brandenberger}\thanks{rhb@physics.mcgill.ca}
\affiliation{Physics Department, McGill University, 3600 University Street, Montreal, QC, H3A 2T8, Canada}

\begin{abstract}
The search for cosmic strings has been of renewed interest with the advent of precision cosmology. In this 
note we give a quantitative description of the nonlinear matter density fluctuations that can form from a
scaling network of cosmic string wakes. Specifically, we compute the distribution of dark matter halos. 
These halos would possess strong correlations in position space that should have survived until today. 
We also discuss the challenges involved in their detection due to their small size and the complex 
dynamics of their formation.
\end{abstract}

\maketitle

\section{Introduction}

Cosmic strings are linear topological defects predicted by many models of particle physics that go
beyond the Standard Model, e.g. in supergravity \cite{jeannerot1}, brane inflation \cite{sarangi1} 
and "String Gas Cosmology" \cite{rhb1989} models. In models which admit these defects, a 
network of cosmic strings is inevitably \cite{Kibble} formed during a phase transition in the very 
early universe and will persist to the present time. This is true as long as the phase transition occurred 
after the period of inflation (provided there was inflation). Since strings carry energy, they lead to
gravitational effects which induce distinctive signatures in cosmological observations. Since the 
string tension (which equals the mass per unit length $\mu$) increases as the energy scale $\eta$ at the time
of string formation, the magnitude of the predicted signatures of strings increases as the
energy scale grows. Thus, a search for observational signatures of strings allows us to constrain 
the physics occurring at energies much larger that can be probed by earth-based particle accelerators.

It has long been known that cosmic strings are able to produce many interesting effects for cosmology. 
In particular, cosmic strings lead to a scale-invariant spectrum of cosmological perturbations
\cite{early}.  Initially, the focus of interest was on strings which have large enough tension to explain 
the entire amplitude of the power spectrum \cite{early}. Such strings would need to have
a value of the tension given (in dimensionless units) by $G \mu \sim 10^{-6}$, where 
$G$ is Newton's gravitational constant. The perturbations produced by cosmic strings
are, however, active and incoherent \cite{active} and hence do not lead to acoustic oscillations
in the angular power spectrum of cosmic microwave (CMB) anisotropies. Thus, when conclusive
evidence for the existence of these oscillations was reached \cite{Boomerang}, it became clear that 
cosmic strings could not be the dominant source of structure formation \cite{perivo1} and 
interest in cosmological effects of strings collapsed. However, in light of the fact that in many models of
inflation (or other explanations for the dominant Gaussian component of the power spectrum of
cosmological perturbations) cosmic strings are predicted, there has lately been a revival of
interest in searching for cosmic strings (see e.g. \cite{RHBrev} for recent surveys). A
second reason for the revival of interest in searching for strings is that there have been great
advances and technological breakthroughs in observational cosmology which are now making it
possible to search for signals of cosmic strings with much lower tensions than those
previously considered.

The searches have been progressing in many directions. First of all, cosmic strings will
contribute to the power spectrum of density fluctuations \cite{early} and to associated
CMB anisotropies \cite{CMBanis}. Signatures of these fluctuations can be searched
for in optical and infrared galaxy surveys and in CMB temperature anisotropy maps. 
Cosmic strings leave behind distinctive signals in 21cm redshift surveys \cite{Holder2}
and in CMB polarization maps \cite{Holder1}. Considerable work has been done to study
the effects of strings in the angular power spectrum of CMB temperature and polarization 
maps \cite{pogosian1}. Analyses combining the SPT \cite{SPT} and WMAP 7-year \cite{WMAP} 
data were able to place the bound $G\mu\leq 1.7\times 10^{-7}$ \cite{dvorkin1} 
\footnote{See \cite{CSbound} for earlier bounds using only the WMAP data.}. Similar bounds are obtained \cite{Bevis} using data from the ACT telescope \cite{ACT} and by the recent Planck survey \cite{planck10}. 
Improved bounds might be achievable by analyzing CMB maps using statistical
tools which are designed to pick out the string-induced non-Gaussianities (see e.g.
\cite{Danos, Sheth} for some recent studies).

In addition to these purely gravitational
effects, cosmic strings can be responsible for the production of highly energetic bursts of particles
 \cite{berezinsky2}, electromagnetic radiation in a wide range of frequencies \cite{cai1} and they
can help seed coherent magnetic fields on galactic scales \cite{rhb1999}. Cosmic
string loops decay by emitting gravitational radiation \cite{Tanmay}. Decay into particle is also possible but the efficiency of particle production from strings remains uncertain. Studies using Nambu-Goto strings and fields theory strings yield different results. Studies using Nambu-Goto strings illustrate that non-superconducting strings decay predominantly via emitting gravitational radiation, whereas field theory simulations indicate that the decay into particles can also be important  (see e.g. \cite{partprod1}).
In any case, strings lead to a scale-invariant stochastic background of gravitational waves
\cite{CSgravwaves} which can be constrained by pulsar timing measurements.
Cusps on string loops may lead to additional emission of gravitational waves \cite{damour1},
and a resulting constraint on the cosmic string tension of 
$G\mu\leq 4\times 10^{-9}$ \cite{haasteren1} has been reported. However the bound 
is sensitive to details of cosmic string cusps which are subject to potentially large
back-reaction effects \cite{CSBReffects}. A bound of $G\mu\leq 5.2\times 10^{-7}$ 
resulting from constraints on the amplitude of gravitational radiation from pulsar
timing constraints is more realistic \cite{sanidas1}. 

Observational optical astronomy is also experiencing a phase of rapid progress.
Larger telescopes are probing the universe at increasing depth, i.e. at increasing
redshift. Since cosmic strings produce non-linear density perturbations at
arbitrarily large redshifts, it is expected that the signals of cosmic strings will stand
out from the structures produced by the main source of fluctuations (Gaussian
fluctuations) more clearly at high redshifts than at low ones. Signals from structures 
seeded by string loops have been studied in a recent paper by Schlaer et al. \cite{schlaer1} 
(see also \cite{Moessner} for earlier work) who found that loops could cause significant star 
formation to occur at high redshift. 
Therefore cosmic strings with enough mass will have an impact on the epoch of 
reionization \cite{Tashiro1, olum2}. String loops would also seed dense dark
matter clumps.  If dark matter self-annihilates, the Fermi
telescope may be sensitive to gamma-rays from the population of these 
clumps today \cite{berezinsky1}. 

In this note we address early structure formation from cosmic string wakes \cite{deruellewakes}. Wakes
lead to cosmic structures with a distinctive shape in position space maps. These
non-Gaussian features may allow the signals of strings to be detected in a
background of Gaussian noise even if the amplitude of the string-induced
signal is quite low, as studied for CMB anisotropy maps in \cite{Danos}.
In Section \ref{background} we give some background about the formation and 
properties of cosmic string wakes. In Section \ref{sec_massfunction} we investigate
the  contribution of string wakes to structure formation as a function of redshift. 
Finally we discuss the implications and observational signals in Section \ref{discussion}.

\section{Background} \label{background}

According to the Kibble mechanism \cite{Kibble}, the formation of cosmic strings from 
a symmetry breaking phase transition implies the existence of a string network. This 
network is expected to evolve towards a scaling solution which states that its properties 
become constant if all lengths are scaled to the Hubble radius \cite{kibble2}. The 
network is composed of two parts. The first is a network of ``infinite" strings
(which includes loops with radius greater than the Hubble radius) whose curvature 
radius $R_c$ can be shown to be of the order of the Hubble scale, i.e. $R_c = \gamma t$ 
with $\gamma\sim \mathcal{O}(1)$. The fact that the curvature radius scales with $t$ is
maintained by intersections and self-intersections of the long strings. Such
intersections create string loops that detach from the long string network, thus allowing
the latter to straighten out. This brings us to the second part of the network: a distribution of 
string loops with curvature radius much smaller than the Hubble scale. 
Following \cite{perivo1} we model a long string as a superposition of straight string segments,
each of whose length is given by $R_c$. This is a reasonable model since any 
wiggles on the strings would either be redshifted away or would cause self-intersections.
The scaling solution suggests $\tilde{N} \sim 1$ long string per Hubble volume, which 
is supported by numerical simulations which yield $\tilde{N} \approx 1-10$ with a loop 
distribution possessing a scaling peak at the loop radius of size $\approx 1/20$ of the 
Hubble radius \cite{olum1} (for earlier work on cosmic string simulations using Nambu-Goto strings see e.g. \cite{CSsimuls} and using field theory strings see e.g. \cite{CSsimuls2}). 

The geometry of space around a long straight string will be conical with deficit angle 
$\alpha = 8\pi G\mu$ \cite{deficit}. A long string moving with speed $v_s$ will 
sweep a plane on which residing static observers would see that matter acquires a velocity 
kick $v=4\pi G\mu v_s \gamma_s$ towards the plane (where $\gamma_s$ is
the relativistic gamma factor associated with the velocity $v_s$). The streams of matter will overlap and 
create a wedge with twice the background density \cite{wake}.
The overdense region will eventually collapse under gravitational instability and form a 
virialized planar structure called the string wake. The growth of the wake will proceed by 
matter accretion. We will briefly review the dynamics of this process.

Let us  consider \footnote{Such calculations were first performed for string wakes in
\cite{wakeZel}.} the effect of the wake on a particle's motion by labelling its physical 
distance to the wake by $r=a(t)(x+\psi (t))$, where $x$ is the comoving position, 
$\psi (t)$ the displacement and  $a(t)$ the scale factor. The initial conditions are given by 
$\psi(t_i)=0$ and $\dot\psi(t_i)= - \text{sgn}(x) v$. The Zel'dovich approximation 
\cite{zeldovich1} gives as equation of motion,
\begin{equation}\label{eq_zeldovichwake}
\ddot\psi +\frac{4}{3t}\dot\psi -\frac{2}{3t^2}\psi =0.
\end{equation}
This system can be solved to obtain,
\begin{equation}\label{eq_solutionzelwake}
\psi(t)=-\frac{12}{5} G\mu \pi v_s \gamma_s t_i \Big( \frac{t}{t_i} \Big)^{2/3} + \frac{12}{5} G\mu \pi v_s \gamma_s \frac{t_i^2}{t}. 
\end{equation}
After dropping the decaying mode, we can find the distance at which our particle decouples 
from the Hubble flow through the condition $\dot r =0$ or more specifically $x+2\psi =0$ 
which yields,
\begin{equation}\label{eq_waketurnaround}
x_{ta}(t)=\pm \frac{24}{5} G\mu \pi v_s \gamma_s t_i \Big( \frac{t}{t_i} \Big)^{2/3} \, .
\end{equation}
Hence $w(t)=\frac{1}{2}a(t)x_{ta}(t)$, which denotes the distance from the wake's center of 
the mass shell that turns-around at $t$, is given by
\begin{equation}\label{eq_widthwake}
w(t)=\frac{12}{5} \pi G \mu v_s \gamma_s t_i \Big( \frac{t}{t_i}\Big)^{4/3}.
\end{equation}

Matter that turned around at a height $w(t_{ta})$ will virialize at a radius $\frac{1}{2}w(t_{ta})$ 
showing that $w(t)$ labels the width of the wake which contains four times the background 
density of matter. Shells of infalling baryonic matter will collide with one another at the 
virial height and create shocks on either side of the wake as shown through hydrodynamical 
simulations \cite{rhb1, haramiyoshi}. The energy of the falling particles will be thermalized 
in the wake giving a temperature \cite{Holder2}, 
\begin{equation}\label{temperatureenergy}
\frac{3}{2} k_B T = \frac{1}{2} m v_{shell}^2,
\end{equation}
where $v_{shell}$ is the speed of hydrogen particles of mass $m$ at the shock. A shell that 
turns around at time $t_{ta}$ will hit the wake at $t=(1+1/\sqrt{2})^{3/2}t_{ta}$, and therefore the
speed will be given by
\begin{align}\label{speedofshell} 
v_{shell}&=\dot r (t,t_{ta}=t(1+1/\sqrt{2})^{-3/2})\nonumber\\
&=\frac{4}{5}(3-2\sqrt{2})v\Big(\frac{t}{t_i}\Big)^{1/3},
\end{align}
which yields a wake temperature of
\begin{align}\label{temperature}
T&=\frac{16}{75}(4(3-2\sqrt{2}))^2 \frac{m}{k_B}(G\mu)^2(v_s\gamma_s)^2\Big(\frac{t}{t_i}\Big)^{2/3}\\
&\simeq [10K](G\mu)^2_6 (v_s \gamma_s)^2\frac{z_i+1}{z+1},
\end{align}
with $(G\mu)_6$ being the value of $G\mu$ in units of $10^{-6}$.

\section{The Mass Function}\label{sec_massfunction}

To compute the amount of matter accreted onto wakes, and eventually the mass function, we 
must understand how to model the network of wakes created from our scaling solution. 
As previously mentioned, we expect $\tilde{N}$ strings per Hubble volume. The curvature
radius of the long strings network grows in length relative to fixed comoving
coordinates. This growth is realized because long string segments intersect and
produce string loops. The string intercommutations occur roughly once per Hubble
expansion time per string per Hubble volume. They also lead to a change of the
direction of motion of a string after intersecting another string and exchanging ends. 
We will use a toy model introduced in \cite{perivo1} to characterize the long string
network analytically: we divide the time axis into Hubble time steps. In each time
step and in each Hubble volume we lay down $\tilde{N}$ straight string
segments of length $\gamma t$ where $\gamma$ is less but close to unity, and each string 
has a velocity $v_s$ in a random direction. We take the string segments to be uncorrelated.
Each string segment lives for a Hubble time. The string network at different Hubble
time steps is taken to be uncorrelated. 

Wakes first form behind strings at the time of equal matter-radiation energy density 
$t_{eq}$, when matter perturbations can start to grow. We consider Hubble time steps 
from $t_{eq}$ to $t_0$ labelling them $t_1=t_{eq},~t_2=2 t_{eq},...,t_m=2^{m-1}t_{eq}$. 
At the beginning of each time step $t_m$ we create $\tilde{N}$ string segments, at the end 
of the time step we remove them. Each of these strings will create a wake 
with initial dimensions $\gamma t_m \times v_s \gamma_s t_m \times w(t_m)$ (where $\gamma_s$
is the relativistic gamma factor associated with the string velocity $v_s$) which will 
then grow in planar dimensions due to the Hubble expansion and in width via the
matter accretion process discussed in the previous section.

\begin{figure}[htbp]
\centering
\includegraphics[width=0.51\textwidth]{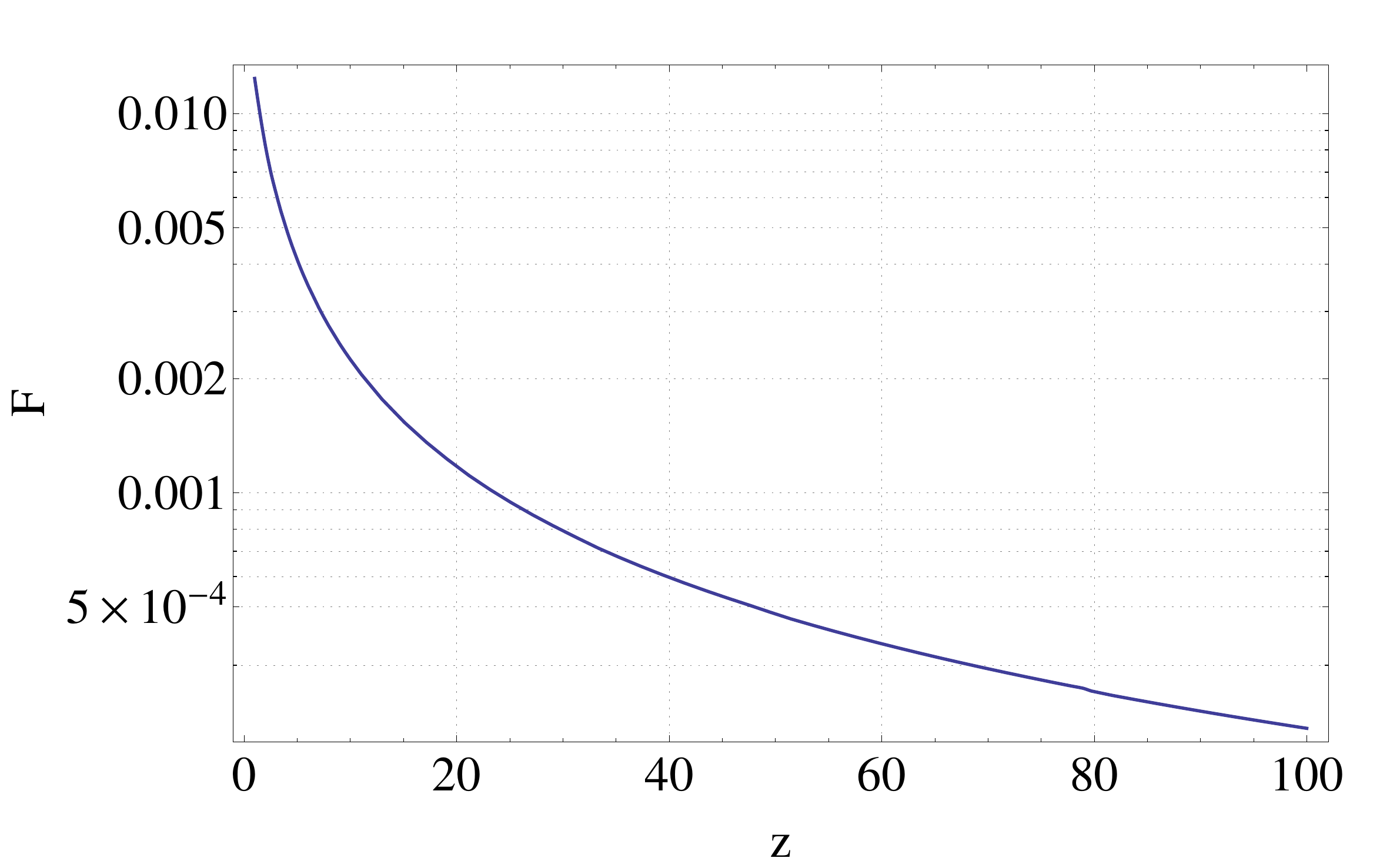}
\caption{\label{massfrac} Fraction of the total matter in the universe that is in cosmic 
string wakes with parameters \ref{parameters} at various redshifts.}
\end{figure}

Consider a redshift z, then the comoving number density of wakes laid down at 
time $z_m>z$ is given by
\begin{equation}\label{eq_wakedens}
\tilde{n}_{wake}(z_m) = \frac{\tilde{N} }{H_m^{-3}} \frac{1}{(z_m+1)^3}.
\end{equation}
Therefore the mass density in all wakes at a redshift $z$ is given by 
summing the contribution from all the wakes created at different Hubble time steps:
\begin{equation}
M_w(z) = 4\rho_b(z) \sum_{z_m > z}{\tilde{n}}_{wake}(z_m) (1+z)^3 \text{Vol}_{wake}(z_m,z) \, ,
\end{equation}
where the last factor is the volume at redshift $z$ of the nonlinear region about a wake created at
a redshift $z_m > z$. Since the planar dimensions of the wake expand with the Hubble
flow, we obtain
\begin{align}\label{Massinwakes}
M_w(z) &=4 \rho_b(z)\sum_{z_m > z}{\tilde{n}}_{wake}(z_m)(1+z)^3 \nonumber \\
&~~~~~~~~~~~~\times (\gamma t_m \times v_s \gamma_s t_m \times w(z))\Big(\frac{z_m+1}{z+1}\Big)^2 \nonumber\\
&= 4\rho_b(z)\sum_{z_m > z}\frac{32}{45}\pi \tilde{N} G\mu \gamma v_s^2\gamma_s^2\frac{1+z_m}{1+z}.
\end{align}
In Figure \ref{massfrac} we plot the quantity $M_w(z)/\rho_b(z)$ (where $\rho_b(z)$ is the
background energy density) for the following parameter choices:
\begin{equation}\label{parameters}
G\mu = 1.5\times 10^{-7},~\tilde{N}=10,~v_s = 1/2~\text{and} ~\gamma =1,
\end{equation}
where the value of $G \mu$ is close to the current upper bound and the other
values are taken to agree with those in recent cosmic string evolution simulations. 
This quantity represents the fraction of the total mass in the universe accreted onto wakes.\\
 
Figure \ref{massfrac} shows that a considerable fraction of matter is accreted by wakes.
Let us analyze this a bit more carefully. Small density perturbations on the surface of 
the wake will grow due to gravitational instability and might eventually cause the wake 
to fragment. The dynamics of this process will be quite complex due to the inner structure 
of the wake: the dark matter density peaks at the edges while baryonic matter clusters 
in the center \cite{rhb1}. An analysis of the fragmentation was done in the simpler 
case of a static isothermal sheet-like cloud in \cite{miyama1}. The timescale of the 
perturbation growth is of the order of the freefall time with the length of the most 
unstable mode equalling the sheet's width and the longest wavelength unstable mode being 
twice that. Taking the very optimistic route and assuming our wake behaves in a similar 
way\footnote{The neglect of Hubble flow along the wake might not be justified. In an 
expanding universe structures with planar, cylindrical and spherical symmetry undergo
a self-similar growth proportional to $t^{4/3},~t~\text{and}~t^{8/9}$ respectively 
\cite{fillmore1}. In this case one could argue that if perturbations grew as cylindrical 
or spherical objects, they would grow slower than the planar wake and therefore no clear 
fragmentation would occur.}, long filaments of virial diameter $\approx w(t)$ should form 
in a similar fashion to the Zel'dovich pancakes.  These filaments would subsequently 
break into beads themselves, again the fastest growing mode having a length 
$\approx 2\pi w(t)$ \cite{schlaer1}. Therefore the wakes will eventually fragment into 
virialized halos of size $\approx \pi (w/2)^3$. 

With this picture in mind we can compute the comoving number density of halos of mass 
greater than $M$ that fragments from the wakes created at redshift $z_m$:
\begin{align}\label{wakehalonum}
   n(z,&z_m,>M)=\tilde{n}_{wake}(z_m) \frac{\text{Vol}_{wake}}{\text{Vol}_{halo}} \Theta (M_{halo}-M)\nonumber\\
&=\tilde{n}_{wake}(z_m) (\gamma t_m \times v_s \gamma_s t_m\times w(z)) \Big( \frac{z_m +1}{z+1} \Big) ^2\nonumber\\
&~~~~~~~~~~\times \frac{1}{\frac{4}{3} \pi w(z)^3}\Theta (2\rho_b(z) \frac{4\pi}{3} w(z)^3-M),
\end{align}
where $\Theta$ is the Heavyside step function.

Taking into account the all wakes that exist at redshift $z$ we simply get 
\begin{equation}
n(z,>M)=\sum_{z_m>z}  n(z,z_m,>M) \, .
\end{equation}
The collection of step functions given by the sum is a bit odd as a distribution since in reality 
this should be continuous. The discrepancy arises because of the model that considers the 
wakes to be instantaneously created while this process actually spans a Hubble time. The 
part of a wake that is created close to time $t_m$ will have different properties than the 
part created at around $2 t_m$ which would be more similar to the beginning of the next set 
of wakes. Hence we should smear out the distribution $n(z,>M)$ to have something realistic. 
To do this we fix $z$ and split the mass range in intervals $\Delta M_m=[M(z,z_m),M(z,z_{m-1})]$ 
with $M(z,z_j)$ being the mass of the halo's formed at $z$ from a wake created at the Hubble 
time step $z_j$. Since the values of $n(z,>M(z_j))=\sum_{z_m>z_j} n(z,z_m,>M)$ are known, 
we can just interpolate to create a continuous $n(z,>M)$. One can then obtain the 
comoving mass function from $\frac{d n(z,>M)}{d\log{M}}$.

\begin{figure}[htbp]
\centering
\includegraphics[width=0.5\textwidth]{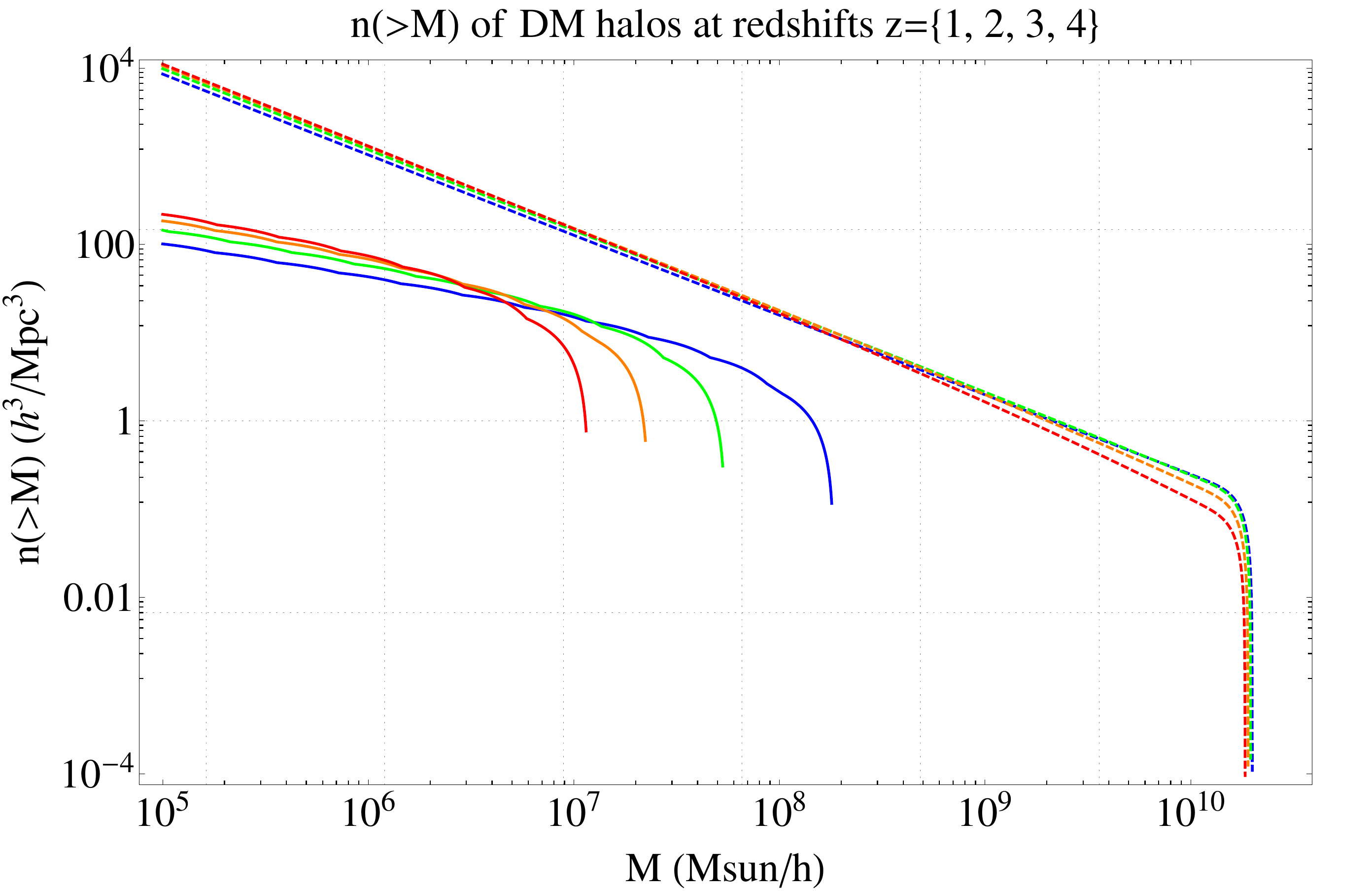}
\includegraphics[width=0.5\textwidth]{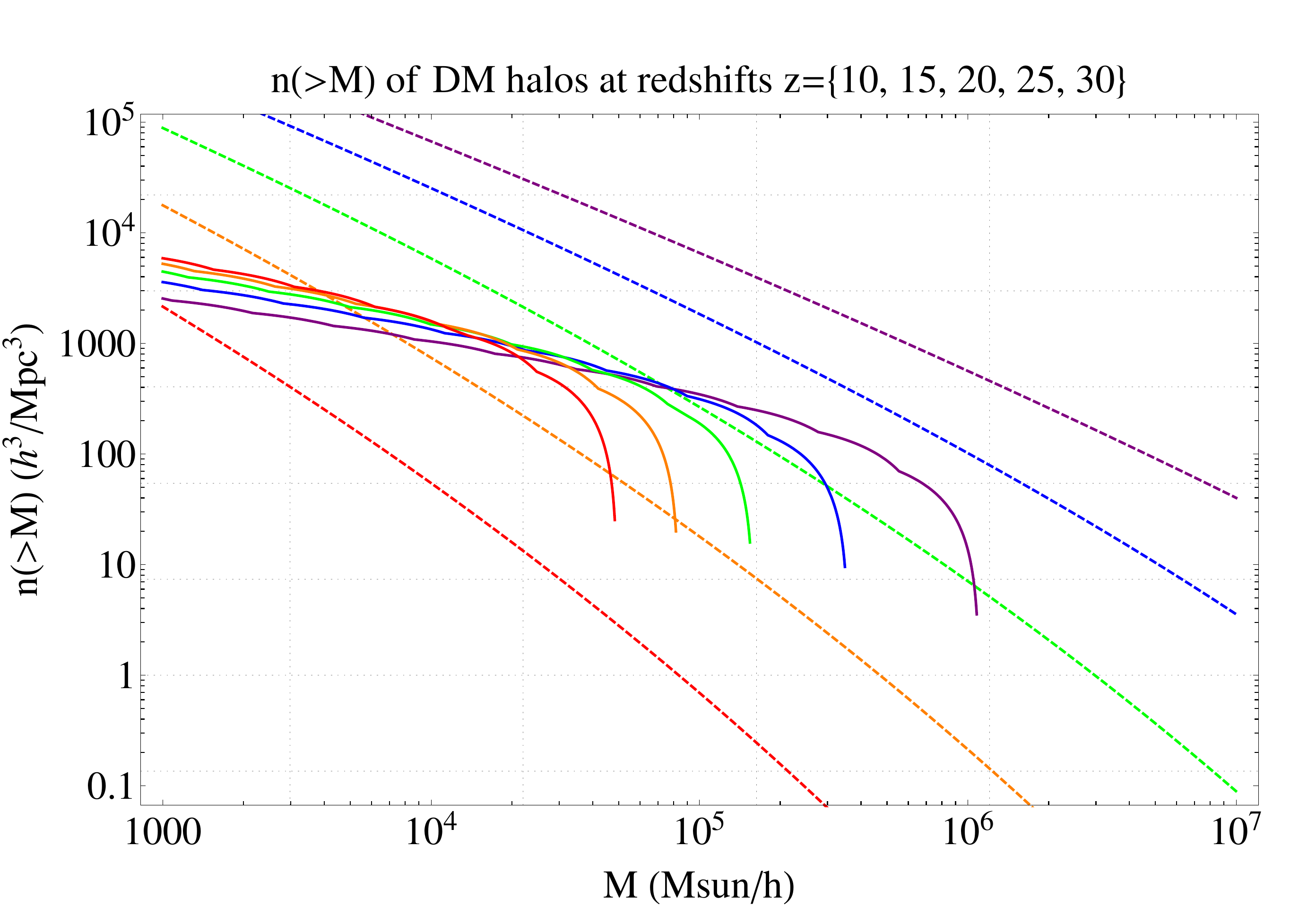}
\caption{\label{figngtM} $n(z,>M)$ of dark matter halos for a string network with parameters \ref{parameters} (solid) and the Reed et al. \cite{reed1} predictions (dashed) at different redshifts. The colors with largest wavelength corresponds to the largest redshifts.}
\end{figure}

\begin{figure}[htbp]
\centering
\includegraphics[width=0.5\textwidth]{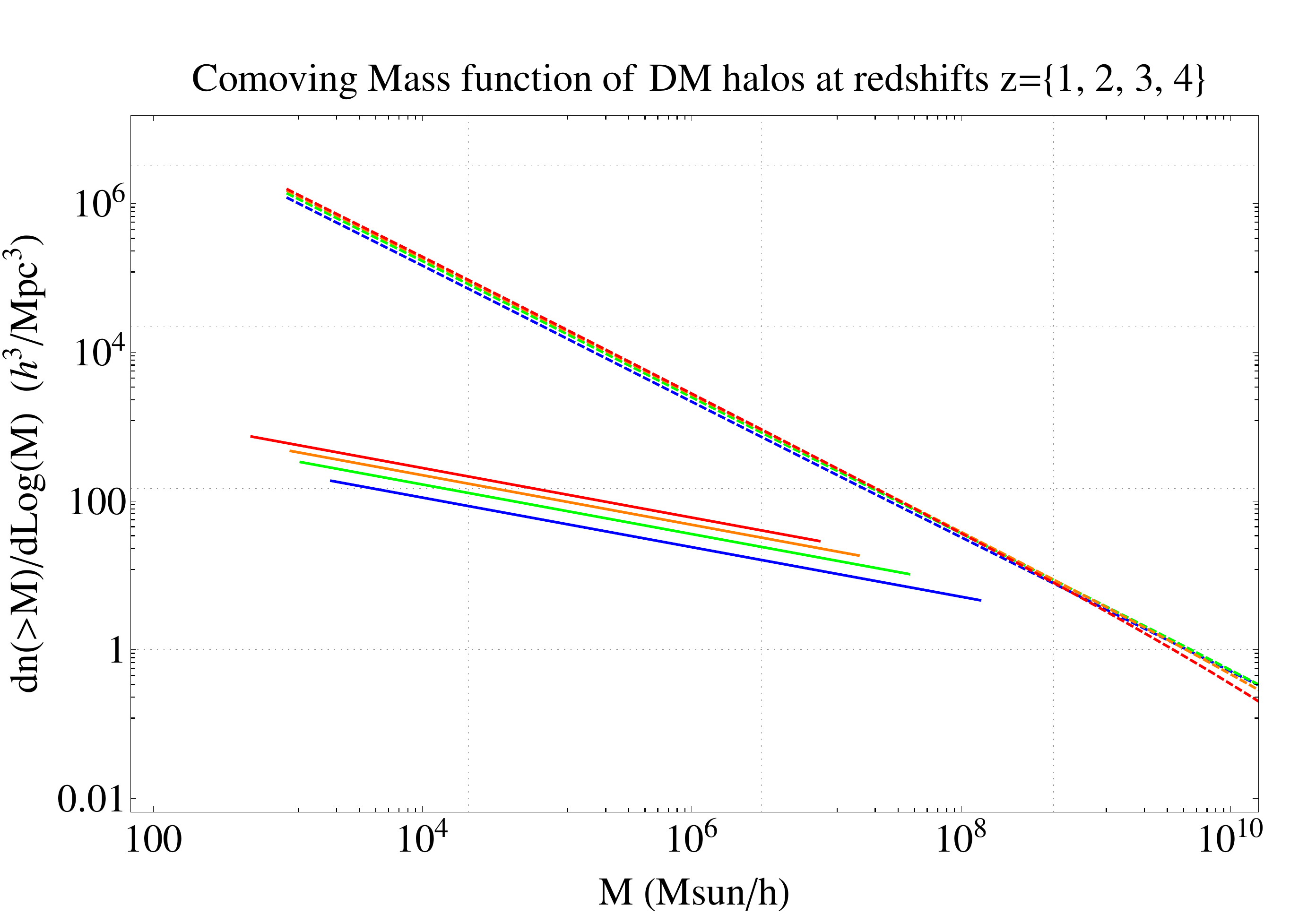}
\includegraphics[width=0.5\textwidth]{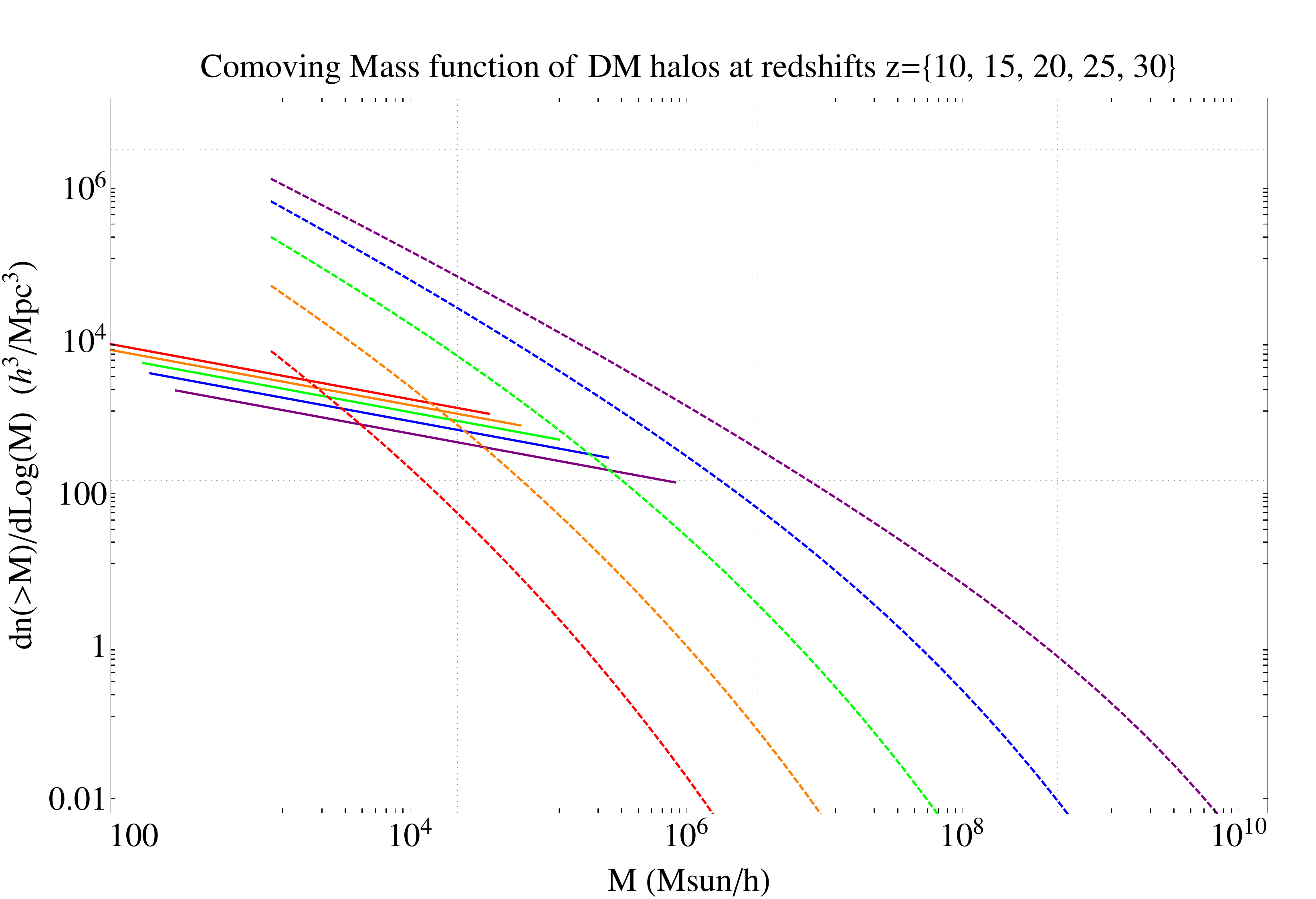}
\caption{\label{fighalomfcomov} Mass function of dark matter halos for a string network with parameters \ref{parameters} (solid) and the Reed et al.\cite{reed1} predictions (dashed) at different redshifts. The colors with largest wavelength corresponds to the largest redshifts.}
\end{figure}

In Figures \ref{figngtM} and  \ref{fighalomfcomov}, we plot and compare $n(z,>M)$ and
$\frac{d n(z,>M)}{d\log{M}}$ from our model to the standard predictions 
as found in Reed et al. \cite{reed1} based on Gaussian fluctuations in a standard
$\Lambda$CDM model with cosmological parameters chosen to agree with those
obtained from the WMAP data. We use the string parameters given in 
Equation \ref{parameters} and note that the small wiggles in the plots of $n(z,>M)$ are 
just an artefact of our interpolation.  We see from the figures that structures formed 
from primordial fluctuations dominate at redshifts smaller than $z\sim 20$, whereas the
string wakes dominate at higher redshifts. One thing to note is that the mass function for 
Gaussian perturbations is obtained using the extended Press-Schechter formalism \cite{bond1} 
which counts the small halos and their parent halos independently. For wakes, we assume  here
that smaller halos get destroyed when they merge in the wake to create larger ones, and 
hence the tail end of the halo distribution could be larger than what we predict
here. 

Note that even if at some redshift the wake halos make up a decent fraction of the total halo 
number in the small mass region, objects of these sizes are hard to detect. Therefore it will be
challenging to probe this mass range directly with any decent precision. 

The statistics studied here do not take into account the distinct position space correlations 
that wake halos would inherit between each other. The wake-induced halos would be
concentrated in sheets, and hence might be visible above a Gaussian noise with significantly
larger amplitude using topological statistics such as Minkowski functionals \cite{Minkowski}.
In the early days of interest in cosmic strings, the application of Minkowski functionals to
distinguish structures seeded by strings from that seeded by Gaussian noise was analyzed in
\cite{early2}. Minkowski functionals have also recently been applied to the analysis of
the signals of cosmic strings in 21cm redshift maps \cite{Evan}. 

\begin{figure}[htbp]
\centering
\includegraphics[width=0.57\textwidth]{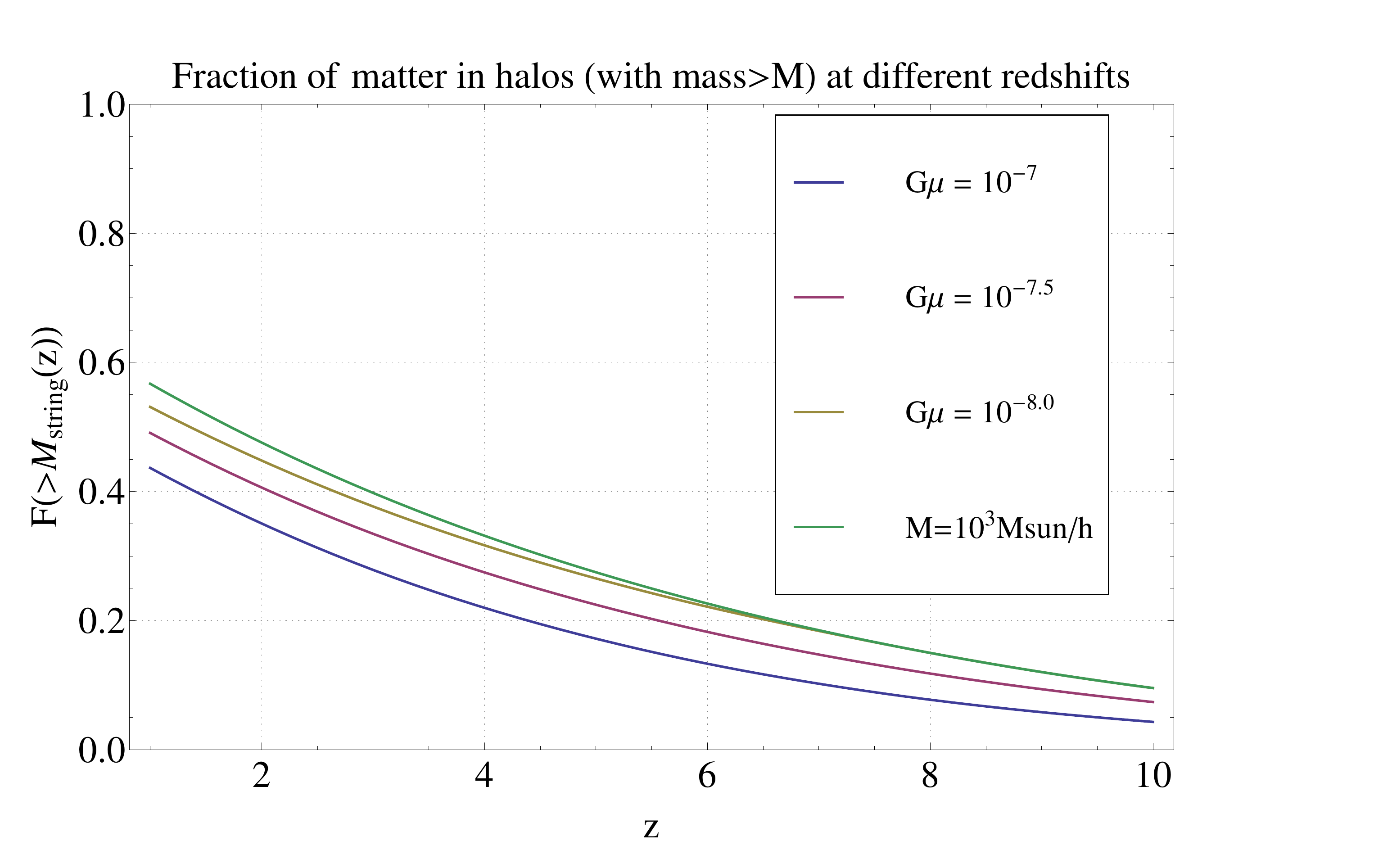}
\caption{\label{destroyallwake} Fraction of matter in halos of mass greater than wakes with parameters \ref{parameters} can create. We plot the curves for multiple values of $G\mu$ and one of a mass of $10^3 \text{M}_\odot$ for reference. These were computed with the program supplied in \cite{reed1}.}
\end{figure}

Note, however, that
the position space correlations might be disrupted by the interactions between the wakes and 
the primordial perturbations, something that we have so far neglected. 
The structures corresponding to density fluctuations on scales larger than the largest wake's 
width cannot  be affected by these interactions. However, the formation of these large objects 
might destroy any wake in their proximity. We give a rough estimate of the number of surviving wakes
by determining the fraction of space that are voids with no large halos. The fraction of space 
that collapsed in Gaussian noise-induced halos of mass greater than $M$ is given 
by \cite{barkanaloeb1},
\begin{align}\label{collapsefraction}
F( > M,z)&=2\int_{\delta_{crit}(z)}^\infty \text{d}\delta \frac{1}{\sqrt{2 \pi }\sigma (M)} \text{exp}\Big(- \frac{\delta^2}{2\sigma(M)^2} \Big) \nonumber\\
&= \text{Erfc} \Big( \frac{\delta_{crit}(z)}{\sqrt{2}\sigma (M)}\Big).
\end{align}
Here, $\delta_{crit}=1.686 (1+z)$ and $\sigma (M)$ is the $rms$ density fluctuation on 
a comoving length scale $R=(3 M/4\pi\rho_b(0))^{1/3}$ which depends on the power 
spectrum of matter. Figure \ref{destroyallwake} shows $F(>M_{w},z)$ for three different 
values of $G\mu$. Here $M_{w}(z,G\mu)$ is the maximum mass of wake halos. 
We also plot $F(>10^3M_\odot /h,z)$ which was the smallest mass that was resolved in the simulation 
by Reed et al. \cite{reed1}. If $F$ is close to 1, then most of the matter was accreted 
onto large Gaussian noise-induced halos, and therefore large halos wiped out any 
geometrical structures and spatial correlations that wake halos could possess.
As the figure shows, the larger the redshift is, the less will be the washout of string-induced
structures by the Gaussian noise.

\section{Discussion} \label{discussion}

In this paper we have studied the formation of halos of dark matter by cosmic string
wakes as a function of redshift. We have found that the relative importance of string
wakes increases as the redshift increases. However, there are challenges in trying
to detect these signals, and this is the topic we discuss in this section
without giving too many details. 

\begin{figure}[htbp]
\centering
\includegraphics[width=0.45\textwidth]{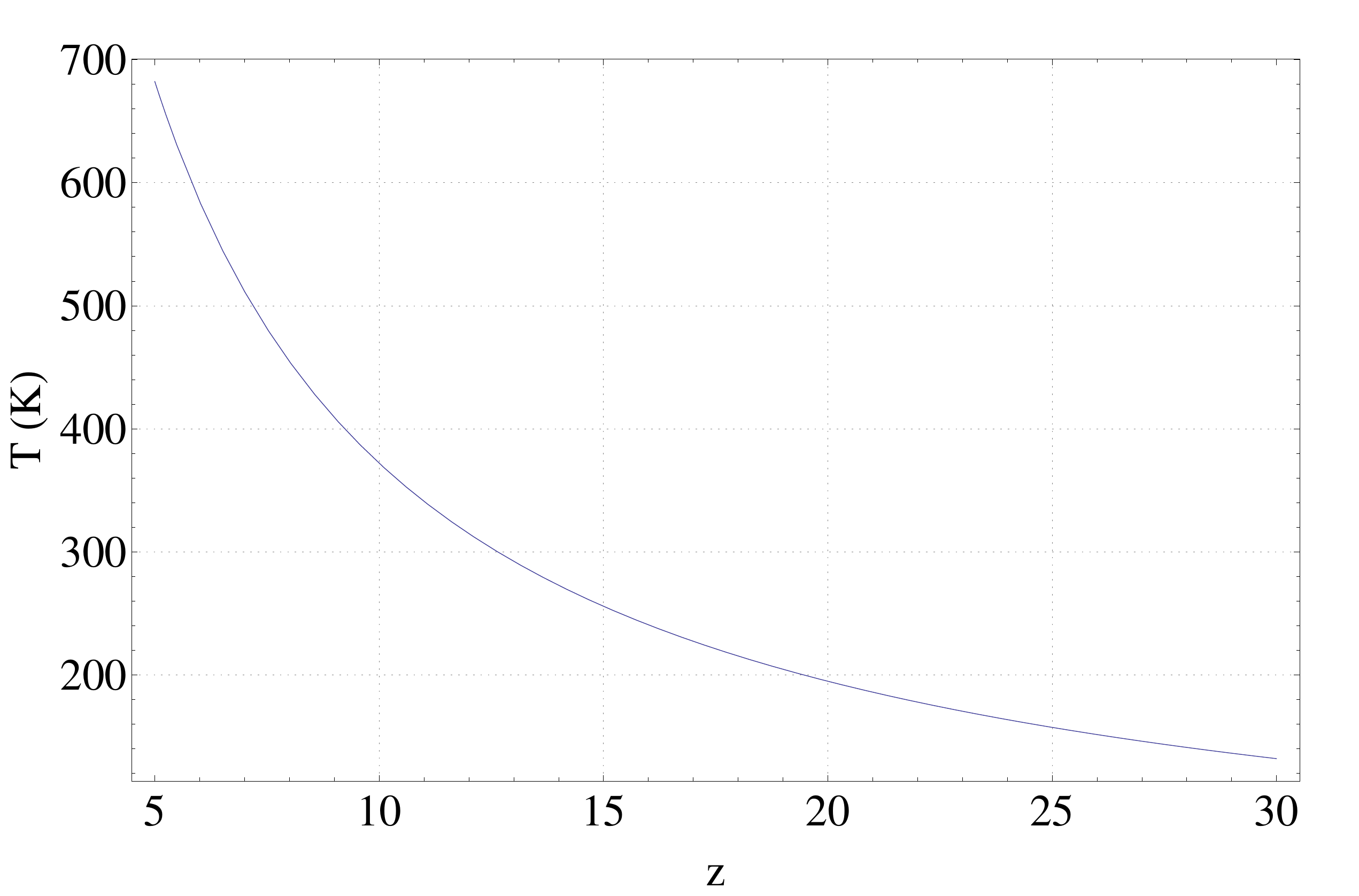}
\caption{\label{fighalotemp} Redshift dependance of the temperature in the hottest halos that are formed 
from wakes. We use the parameters in Eq. \ref{parameters} and $\mu\approx 1.22$. Note that in the
entire redshift range shown, the halo temperature is higher than both the CMB background temperature
and the average kinetic temperature of the baryonic gas.}
\end{figure}

The temperature of our halos can be estimated from the virial theorem and conservation 
of energy by considering a collapsing gas cloud of radius $w$ containing two times the 
background density $\rho_b$,
\begin{align}
2 \frac{1}{2}v_{virial}^2&=\frac{3}{5}\frac{G M}{r_{virial}}, \label{virialtheorem}\\
\frac{1}{2}v_{virial}^2-\frac{3}{5}\frac{G M}{r_{virial}}&=\frac{1}{2}v_{th}^2-\frac{3}{5}\frac{G M}{w} \, .\label{energyconservation}
\end{align}
Here, $w$ is the radius of our collapsing cloud from eq. \ref{eq_widthwake}, 
$M=2\rho_b \frac{4}{3}\pi (w)^3$ is its mass and $v_{th}=v_{shell}$ is the initial thermal 
velocity in Eq. \ref{speedofshell}. The shock heated baryonic gas will have  a temperature of 
\begin{equation}
T = \frac{\mu}{2} \frac{m_{p}}{k_{b}}v^2,
\end{equation}
where $m_{H}$ is the mass of the proton 
and $\mu$ is the mean molecular weight \cite{barkanaloeb1}. Star formation can occur 
if the temperatures are high enough so that colliding molecules can excite each other to 
higher energy states and radiate away their kinetic energy. Atomic cooling requires 
$T>10^4$K while $H2$ cooling has channels that can be efficient down to temperatures as 
low as $200$K. However these $H2$ molecules are very fragile and get destroyed by the 
radiation of the first stars, and therefore this cooling mechanism alone will not allow for 
multiple stars to form in close proximity as would be the case for halos formed from wakes. 
Figure \ref{fighalotemp} shows the temperature of the hottest temperatures that can be 
achieved with the parameters in Eq. \ref{parameters} and $\mu\approx 1.22$ in the case of 
neutral hydrogen. As we can see, the temperatures are far too low to have any significant 
star formation, and hence no noticeable effect on the history of reionization will be present.

Gravitational lensing could also be an interesting window considering there should 
be about $\sim10^6$ wakes in our Hubble volume. However, even if each wake contains 
a lot of mass, it is spread over a large area and hence we would only obtain strong lensing 
if the background source lay, with us, very close to the plane delimited by a wake, in order to 
have enough mass along the integrated line of sight. This makes the signal too rare to be useful 
to statistically constrain string parameters. However, if one were lucky, assuming the images 
do not overlap, which could happen for extended source and a small $G\mu$ or if we are not in 
the proximity of the wake's plane, a lensed source would appear as two images with no 
magnification or distortion. In such case one could hope to find other similarily lensed objects 
along some line in the sky. 

Observables sensitive to the structures formed at high redshift would be ideal to differentiate 
and detect wake structures from the background. One such example, that relies on the 
compact objects, is gamma ray constraints from dense dark matter clumps 
\cite{berezinsky1, pat1}.  However, the formation of these objects must proceed by a 
clean radial infall of matter in order to have a sufficiently steep dark matter 
density profile to produce a detectable gamma-ray flux. 
These are not favorable conditions in the wake since we expect strong tidal forces as 
growing perturbations 
oscillate around the center.

In conclusion, we have shown that string wakes can dominate the nonlinear structures in the
universe at sufficiently high redshifts. Even at redshifts where they are not dominant, they
may be identifiable because of the distinct spatial correlations which they induce. 
However, the string-induced halos are typically too small and cold to induce star
formation, and hence it is a challenge to be able to find them observationally. One
promising window - studied in other works \cite{Holder2, Wangyi, Oscar, Pagano, Evan} - is via 
21cm redshift surveys.

\section{Acknowledgements}
We would like to thank P. Scott, G. Holder, Y. Omori, Y. Hezaveh, J. Harnois-Deraps and U.L. Pen for useful discussions 
and comments. This work has been supported in part by an NSERC Discovery Grant and by funds
from the Canada Research Chair program.


\begin{thebibliography}{9}

\bibitem{jeannerot1}
R.~Jeannerot,
  ``A Supersymmetric SO(10) Model with Inflation and Cosmic Strings,''
  Phys.\ Rev.\  D {\bf 53}, 5426 (1996)
  [arXiv:hep-ph/9509365];\\
R.~Jeannerot, J.~Rocher and M.~Sakellariadou,
  ``How generic is cosmic string formation in SUSY GUTs,''
  Phys.\ Rev.\  D {\bf 68}, 103514 (2003)
  [arXiv:hep-ph/0308134].

\bibitem{sarangi1}
S.~Sarangi and S.~H.~H.~Tye,
  ``Cosmic string production towards the end of brane inflation,''
  Phys.\ Lett.\  B {\bf 536}, 185 (2002)
  [arXiv:hep-th/0204074];\\
E.~J.~Copeland, R.~C.~Myers and J.~Polchinski,
  ``Cosmic F- and D-strings,''
  JHEP {\bf 0406}, 013 (2004)
  [arXiv:hep-th/0312067].

\bibitem{rhb1989}
R.~H.~Brandenberger and C.~Vafa,
  ``Superstrings in the Early Universe,''
  Nucl.\ Phys.\  B {\bf 316}, 391 (1989);\\
 A.~Nayeri, R.~H.~Brandenberger and C.~Vafa, 
  ``Producing a scale-invariant spectrum of perturbations in a Hagedorn phase 
  of string cosmology,''
 Phys.\ Rev.\ Lett.\  {\bf 97}, 021302 (2006)   [arXiv:hep-th/0511140];\\
 R.~H.~Brandenberger, A.~Nayeri, S.~P.~Patil and C.~Vafa,
  ``String gas cosmology and structure formation,''
 Int.\ J.\ Mod.\ Phys.\ A {\bf 22}, 3621 (2007)
  [hep-th/0608121];\\
  R.~H.~Brandenberger,
  ``String Gas Cosmology,''
  arXiv:0808.0746 [hep-th].

\bibitem{Kibble}
T.~W.~B.~Kibble,
  ``Topology of Cosmic Domains and Strings,''
  J.\ Phys.\ A {\bf 9}, 1387 (1976);\\
T.~W.~B.~Kibble,
  ``Some Implications of a Cosmological Phase Transition,''
  Phys.\ Rept.\  {\bf 67}, 183 (1980);\\
T.~W.~B.~Kibble,
  ``Phase Transitions In The Early Universe,''
  Acta Phys.\ Polon.\ B {\bf 13}, 723 (1982).

\bibitem{early}
 Y.~B.~Zeldovich,
  ``Cosmological fluctuations produced near a singularity,''
  Mon.\ Not.\ Roy.\ Astron.\ Soc.\  {\bf 192}, 663 (1980);\\
 A.~Vilenkin,
  ``Cosmological Density Fluctuations Produced By Vacuum Strings,''
  Phys.\ Rev.\ Lett.\  {\bf 46}, 1169 (1981)
  [Erratum-ibid.\  {\bf 46}, 1496 (1981)];\\
N.~Turok and R.~H.~Brandenberger,
  ``Cosmic Strings And The Formation Of Galaxies And Clusters Of Galaxies,''
  Phys.\ Rev.\ D {\bf 33}, 2175 (1986);\\
H. Sato, ``Galaxy Formation by Cosmic Strings,''
  Prog. Theor. Phys.\  {\bf 75}, 1342 (1986);\\
A. Stebbins, ``Cosmic Strings and Cold Matter'',
  Ap. J. (Lett.) {\bf 303}, L21 (1986).\\
N.~Deruelle,~D.~Langlois,~J.P.~Uzan, "Cosmological Perturbations seeded by Topological Defects: Setting the Initial Conditions", Phys. Rev. D56 (1997) 7608, [arXiv:gr-qc/9707035]


\bibitem{active}
J.~Magueijo, A.~Albrecht, D.~Coulson and P.~Ferreira,
  ``Doppler peaks from active perturbations,''
  Phys.\ Rev.\ Lett.\  {\bf 76}, 2617 (1996)
  [arXiv:astro-ph/9511042];\\
U.~L.~Pen, U.~Seljak and N.~Turok,
  ``Power spectra in global defect theories of cosmic structure formation,''
  Phys.\ Rev.\ Lett.\  {\bf 79}, 1611 (1997)
  [arXiv:astro-ph/9704165];\\
L.~Perivolaropoulos,
  ``Spectral Analysis Of Microwave Background Perturbations Induced By Cosmic
  Strings,''
  Astrophys.\ J.\  {\bf 451}, 429 (1995)
  [arXiv:astro-ph/9402024].
  
\bibitem{Boomerang}
 P.~D.~Mauskopf {\it et al.}  [Boomerang Collaboration],
  ``Measurement of a Peak in the Cosmic Microwave Background Power Spectrum
  from the North American test flight of BOOMERANG,''
  Astrophys.\ J.\  {\bf 536}, L59 (2000)
  [arXiv:astro-ph/9911444].

\bibitem{perivo1}
 L.~Perivolaropoulos,
  ``COBE versus cosmic strings: An Analytical model,''
  Phys.\ Lett.\  B {\bf 298}, 305 (1993)
  [arXiv:hep-ph/9208247];\\
L.~Perivolaropoulos,
  ``Statistics of microwave fluctuations induced by topological defects,''
  Phys.\ Rev.\  D {\bf 48}, 1530 (1993)
  [arXiv:hep-ph/9212228].\\
A.~Riazuelo,~N.~Deruelle,~P.~Peter, "Topological Defects and CMB anisotropies : Are the predictions reliable ?", Phys. Rev. D61. (2000) 123504, [astro-ph/9910290]
 
\bibitem{RHBrev}
R.~H.~Brandenberger,
  ``Topological defects and structure formation,''
  Int.\ J.\ Mod.\ Phys.\ A {\bf 9}, 2117 (1994)
  [arXiv:astro-ph/9310041].\\
E.J. Copeland, L. Pogosian, T. Vachaspati,
  ``Seeking String Theory in the Cosmos,''
  [arXiv:1105.0207].\\
M. Hindmarsh,
  ``Signals of Inflationary Models with Cosmic Strings,''
  [arXiv:1106.0391].\\
R.~H.~Brandenberger,
  ``Searching for Cosmic Strings in New Observational Windows,''
  [arXiv:1301.2856].

\bibitem{CMBanis}
J.~H.~Traschen, N.~Turok and R.~H.~Brandenberger,
  ``Microwave Anisotropies from Cosmic Strings,''
  Phys.\ Rev.\  D {\bf 34}, 919 (1986);\\
R.~H.~Brandenberger and N.~Turok,
  ``Fluctuations From Cosmic Strings And The Microwave Background,''
  Phys.\ Rev.\ D {\bf 33}, 2182 (1986).

\bibitem{Holder2}
 R.~H.~Brandenberger, R.~J.~Danos, O.~F.~Hernandez and G.~P.~Holder,
  ``The 21 cm Signature of Cosmic String Wakes,''
  JCAP {\bf 1012}, 028 (2010)
  [arXiv:1006.2514 [astro-ph.CO]].
  
\bibitem{Holder1}
 R.~J.~Danos, R.~H.~Brandenberger and G.~Holder,
  ``A Signature of Cosmic Strings Wakes in the CMB Polarization,''
  Phys.\ Rev.\  D {\bf 82}, 023513 (2010)
  [arXiv:1003.0905 [astro-ph.CO]].

\bibitem{pogosian1}
L. Pogosian \& M. Wyman, \emph{B-modes from Cos-
mic Strings,} Phys. Rev. D 77, 083509 (2008)
[arXiv: 0711.0747].\\
 A. Avgoustidis et al., \emph{Constraints on the fundamental
string coupling from B-mode experiments,} Phys. Rev.
Lett. 107, 121301 (2011) [arXiv:1105.6198].

\bibitem{SPT}
 J.~E.~Ruhl {\it et al.}  [The SPT Collaboration],
  ``The South Pole Telescope,''
  Proc.\ SPIE Int.\ Soc.\ Opt.\ Eng.\  {\bf 5498}, 11 (2004)
  [arXiv:astro-ph/0411122];\\
  J.~E.~Carlstrom {\it et al.},
  ``The 10 Meter South Pole Telescope,''
  Publ.\ Astron.\ Soc.\ Pac.\  {\bf 123}, 568 (2011)
  [arXiv:0907.4445 [astro-ph.IM]].

\bibitem{WMAP}
D.~Larson, J.~Dunkley, G.~Hinshaw, E.~Komatsu, M.~R.~Nolta, C.~L.~Bennett, B.~Gold and M.~Halpern {\it et al.},
  ``Seven-Year Wilkinson Microwave Anisotropy Probe (WMAP) Observations: Power Spectra and WMAP-Derived Parameters,''
  Astrophys.\ J.\ Suppl.\  {\bf 192}, 16 (2011)
  [arXiv:1001.4635 [astro-ph.CO]].

\bibitem{dvorkin1}
  C.~Dvorkin, M.~Wyman and W.~Hu,
  ``Cosmic String constraints from WMAP and the South Pole Telescope,''
  Phys.\ Rev.\ D {\bf 84}, 123519 (2011)
  [arXiv:1109.4947 [astro-ph.CO]].

\bibitem{CSbound}
L.~Pogosian, S.~H.~H.~Tye, I.~Wasserman and M.~Wyman,
  ``Observational constraints on cosmic string production during brane
  inflation,''
  Phys.\ Rev.\  D {\bf 68}, 023506 (2003)
  [Erratum-ibid.\  D {\bf 73}, 089904 (2006)]
  [arXiv:hep-th/0304188];\\
M.~Wyman, L.~Pogosian and I.~Wasserman,
  ``Bounds on cosmic strings from WMAP and SDSS,''
  Phys.\ Rev.\  D {\bf 72}, 023513 (2005)
  [Erratum-ibid.\  D {\bf 73}, 089905 (2006)]
  [arXiv:astro-ph/0503364];\\
A.~A.~Fraisse,
  ``Limits on Defects Formation and Hybrid Inflationary Models with
  Three-Year WMAP Observations,''
  JCAP {\bf 0703}, 008 (2007)
  [arXiv:astro-ph/0603589];\\
U.~Seljak, A.~Slosar and P.~McDonald,
  ``Cosmological parameters from combining the Lyman-alpha forest with CMB,
  galaxy clustering and SN constraints,''
  JCAP {\bf 0610}, 014 (2006)
  [arXiv:astro-ph/0604335];\\
  R.~A.~Battye, B.~Garbrecht and A.~Moss,
  ``Constraints on supersymmetric models of hybrid inflation,''
  JCAP {\bf 0609}, 007 (2006)
  [arXiv:astro-ph/0607339];\\
R.~A.~Battye, B.~Garbrecht, A.~Moss and H.~Stoica,
  ``Constraints on Brane Inflation and Cosmic Strings,''
  JCAP {\bf 0801}, 020 (2008)
  [arXiv:0710.1541 [astro-ph]];\\
N.~Bevis, M.~Hindmarsh, M.~Kunz and J.~Urrestilla,
  ``Fitting CMB data with cosmic strings and inflation,''
 Phys.\ Rev.\ Lett.\  {\bf 100}, 021301 (2008)
  [astro-ph/0702223];\\
N.~Bevis, M.~Hindmarsh, M.~Kunz and J.~Urrestilla,
  ``CMB power spectrum contribution from cosmic strings using  field-evolution
  simulations of the Abelian Higgs model,''
  Phys.\ Rev.\  D {\bf 75}, 065015 (2007)
  [arXiv:astro-ph/0605018];\\
R.~Battye and A.~Moss,
  ``Updated constraints on the cosmic string tension,''
  Phys.\ Rev.\ D {\bf 82}, 023521 (2010)
  [arXiv:1005.0479 [astro-ph.CO]].
  
\bibitem{Bevis}
J. Urrestilla, N. Bevis, M. Hindmarsh, and M. Kunz,
``Cosmic string parameter constraints and model analysis using small scale Cosmic Microwave Background data,''
  JCAP {\bf 1112}, 021 (2011)
  [arXiv:1108.2730 [astro-ph.CO]].\\
J. Dunkley et al., Astrophys. J. 739, 52 (2011)

\bibitem{ACT}
A.~Kosowsky  [the ACT Collaboration],
  ``The Atacama Cosmology Telescope Project: A Progress Report,''
  New Astron.\ Rev.\  {\bf 50}, 969 (2006)
  [arXiv:astro-ph/0608549].
\bibitem{planck10}
P. A. R. Ade et al. [ Planck Collaboration], ``Planck 2013 results. XXV. Searches for cosmic strings and other
topological defects,” arXiv:1303.5085 [astro-ph.CO].
  
\bibitem{Danos}
S.~Amsel, J.~Berger and R.~H.~Brandenberger,
  ``Detecting Cosmic Strings in the CMB with the Canny Algorithm,''
  JCAP {\bf 0804}, 015 (2008)
  [arXiv:0709.0982 [astro-ph]];\\
A.~Stewart and R.~Brandenberger,
  ``Edge Detection, Cosmic Strings and the South Pole Telescope,''
 JCAP {\bf 0902}, 009 (2009)
  [arXiv:0809.0865 [astro-ph]];\\
  R.~J.~Danos and R.~H.~Brandenberger,
  ``Canny Algorithm, Cosmic Strings and the Cosmic Microwave Background,''
  Int.\ J.\ Mod.\ Phys.\ D {\bf 19}, 183 (2010)
  [arXiv:0811.2004 [astro-ph]].
  
\bibitem{Sheth}
M.~S.~Movahed, B.~Javanmardi and R.~K.~Sheth,
  ``Peak-peak correlations in the cosmic background radiation from cosmic strings,''
  arXiv:1212.0964 [astro-ph.CO].
  
\bibitem{berezinsky2}
V.~Berezinsky, K.~D.~Olum, E.~Sabancilar and A.~Vilenkin,
  ``UHE neutrinos from superconducting cosmic strings,''
  Phys.\ Rev.\ D {\bf 80}, 023014 (2009)
  [arXiv:0901.0527 [astro-ph.HE]];\\
C.~Lunardini and E.~Sabancilar,
  ``Cosmic Strings as Emitters of Extremely High Energy Neutrinos,''
  Phys.\ Rev.\ D {\bf 86}, 085008 (2012)
  [arXiv:1206.2924 [astro-ph.CO]].

\bibitem{cai1}
Y.~-F.~Cai, E.~Sabancilar, D.~A.~Steer and T.~Vachaspati,
  ``Radio Broadcasts from Superconducting Strings,''
  Phys.\ Rev.\ D {\bf 86}, 043521 (2012)
  [arXiv:1205.3170 [astro-ph.CO]];\\
 T.~Vachaspati,
  ``Cosmic Sparks from Superconducting Strings,''
  Phys.\ Rev.\ Lett.\  {\bf 101}, 141301 (2008)
  [arXiv:0802.0711 [astro-ph]];\\
Y.~-F.~Cai, E.~Sabancilar and T.~Vachaspati,
  ``Radio bursts from superconducting strings,''
  Phys.\ Rev.\ D {\bf 85}, 023530 (2012)
  [arXiv:1110.1631 [astro-ph.CO]];\\
V.~Berezinsky, B.~Hnatyk and A.~Vilenkin,
  ``Gamma-ray bursts from superconducting cosmic strings,''
  Phys.\ Rev.\ D {\bf 64}, 043004 (2001)
  [astro-ph/0102366];\\
K.~S.~Cheng, Y.~-W.~Yu and T.~Harko,
  ``High Redshift Gamma-Ray Bursts: Observational Signatures of Superconducting Cosmic Strings?,''
  Phys.\ Rev.\ Lett.\  {\bf 104}, 241102 (2010)
  [arXiv:1005.3427 [astro-ph.HE]].
  
\bibitem{rhb1999}
R.~H.~Brandenberger and X.~-m.~Zhang,
  ``Anomalous global strings and primordial magnetic fields,''
  Phys.\ Rev.\ D {\bf 59}, 081301 (1999)
  [hep-ph/9808306].
  
\bibitem{Tanmay}
T.~Vachaspati and A.~Vilenkin,
  ``Gravitational Radiation from Cosmic Strings,''
  Phys.\ Rev.\ D {\bf 31}, 3052 (1985);\\
R. L. Davis, 
``Nucleosynthesis Problems for String Models of Galaxy Formation",
Phys. Lett. {\bf B 161}, 285 (1985).

\bibitem{partprod1}
S. Borsanyi, M. Hindmarsh, ``Semiclassical decay of topological defects,'' 
Phys.\ Rev.\ D {\bf 77}:045022, (2008), [arXiv:0712.0300].\\
U.F. Wichoski, J.H. MacGibbon, R.H. Brandenberger, ``High Energy Neutrino, Photon and Cosmic Ray Fluxes from VHS Cosmic Strings,'' Phys. Rev. D{\bf 65}, 063005
(2002), [arXiv:hep-ph/9805419].\\
G. R. Vincent, M. Hindmarsh, M. Sakellariadou, ``Scaling and small scale structure in
cosmic string networks,'' Phys. Rev. D{\bf 56} (1997) 637–646, [astro-ph/9612135].\\
G. Vincent, N. D. Antunes, and M. Hindmarsh, ``Numerical Simulations of String Networks in
the Abelian-Higgs Model,'' Phys. Rev. Lett. {\bf 80} ( 1998) 2277–2280, [hep-ph/9708427].
\bibitem{CSgravwaves}
R.~H.~Brandenberger, A.~Albrecht and N.~Turok,
  ``Gravitational Radiation From Cosmic Strings And The Microwave Background,''
  Nucl.\ Phys.\ B {\bf 277}, 605 (1986).
  
\bibitem{damour1}
T.~Damour and A.~Vilenkin,
  ``Gravitational radiation from cosmic (super)strings: Bursts, stochastic background, and observational windows,''
  Phys.\ Rev.\ D {\bf 71}, 063510 (2005)
  [hep-th/0410222];\\
S.~Olmez, V.~Mandic and X.~Siemens,
  ``Gravitational-Wave Stochastic Background from Kinks and Cusps on Cosmic Strings,''
  Phys.\ Rev.\ D {\bf 81}, 104028 (2010)
  [arXiv:1004.0890 [astro-ph.CO]].

\bibitem{haasteren1}
R.~van Haasteren, Y.~Levin, G.~H.~Janssen, K.~Lazaridis, M.~K.~B.~W.~Stappers, G.~Desvignes, M.~B.~Purver and A.~G.~Lyne {\it et al.},
  ``Placing limits on the stochastic gravitational-wave background using European Pulsar Timing Array data,''
  arXiv:1103.0576 [astro-ph.CO].
  
\bibitem{CSBReffects}
R.~H.~Brandenberger,
  ``On The Decay Of Cosmic String Loops,''
  Nucl.\ Phys.\ B {\bf 293}, 812 (1987).
  
\bibitem{sanidas1}
S.~A.~Sanidas, R.~A.~Battye and B.~W.~Stappers,
  ``Constraints on cosmic string tension imposed by the limit on the stochastic gravitational wave background from the European Pulsar Timing Array,''
  Phys.\ Rev.\ D {\bf 85}, 122003 (2012)
  [arXiv:1201.2419 [astro-ph.CO]].
  
\bibitem{schlaer1}
B.~Shlaer, A.~Vilenkin and A.~Loeb,
  ``Early structure formation from cosmic string loops,''
  JCAP {\bf 1205}, 026 (2012)
  [arXiv:1202.1346 [astro-ph.CO]].

\bibitem{Moessner}
R.~Moessner and R.~H.~Brandenberger,
  ``Formation of high redshift objects in a cosmic string theory with hot dark matter,''
  Mon.\ Not.\ Roy.\ Astron.\ Soc.\  {\bf 280}, 797 (1996)
  [astro-ph/9510141].

\bibitem{Tashiro1}
H.~Tashiro, E.~Sabancilar and T.~Vachaspati,
  ``Constraints on Superconducting Cosmic Strings from Early Reionization,''
  Phys.\ Rev.\ D {\bf 85}, 123535 (2012)
  [arXiv:1204.3643 [astro-ph.CO]].

\bibitem{olum2}
K.~D.~Olum and A.~Vilenkin,
  ``Reionization from cosmic string loops,''
  Phys.\ Rev.\ D {\bf 74}, 063516 (2006)
  [astro-ph/0605465].
  
\bibitem{berezinsky1}
V.~S.~Berezinsky, V.~I.~Dokuchaev and Y.~.N.~Eroshenko,
  ``Dense DM clumps seeded by cosmic string loops and DM annihilation,''
  JCAP {\bf 1112}, 007 (2011)
  [arXiv:1107.2751 [astro-ph.HE]].

\bibitem{deruellewakes}
N.~Deruelle and B.~Linet, " A Cosmic String Shock Wave", Class. and Quant. Grav. 5 (1988) 55 

\bibitem{kibble2}
T.~W.~B.~Kibble,
  ``Evolution Of A System Of Cosmic Strings,''
  Nucl.\ Phys.\ B {\bf 252}, 227 (1985)
  [Erratum-ibid.\ B {\bf 261}, 750 (1985)].

\bibitem{olum1}
J.~J.~Blanco-Pillado, K.~D.~Olum and B.~Shlaer,
  ``Large parallel cosmic string simulations: New results on loop production,''
  Phys.\ Rev.\ D {\bf 83}, 083514 (2011)
  [arXiv:1101.5173 [astro-ph.CO]].

\bibitem{CSsimuls}
A.~Albrecht and N.~Turok,
  ``Evolution Of Cosmic Strings,''
  Phys.\ Rev.\ Lett.\  {\bf 54}, 1868 (1985);\\
D.~P.~Bennett and F.~R.~Bouchet,
  ``Evidence For A Scaling Solution In Cosmic String Evolution,''
  Phys.\ Rev.\ Lett.\  {\bf 60}, 257 (1988);\\
B.~Allen and E.~P.~S.~Shellard,
  ``Cosmic String Evolution: A Numerical Simulation,''
  Phys.\ Rev.\ Lett.\  {\bf 64}, 119 (1990);\\
C.~Ringeval, M.~Sakellariadou and F.~Bouchet,
  ``Cosmological evolution of cosmic string loops,''
  JCAP {\bf 0702}, 023 (2007)
  [arXiv:astro-ph/0511646];\\
  A.~A.~Fraisse, C.~Ringeval, D.~N.~Spergel and F.~R.~Bouchet,
  ``Small-Angle CMB Temperature Anisotropies Induced by Cosmic Strings,''
  Phys.\ Rev.\ D {\bf 78}, 043535 (2008)
  [arXiv:0708.1162 [astro-ph]];\\
V.~Vanchurin, K.~D.~Olum and A.~Vilenkin,
  ``Scaling of cosmic string loops,''
  Phys.\ Rev.\  D {\bf 74}, 063527 (2006)
  [arXiv:gr-qc/0511159];\\
  C.~Ringeval and F.~R.~Bouchet,
  ``All sky CMB map from cosmic strings integrated Sachs-Wolfe effect,''
  arXiv:1204.5041 [astro-ph.CO].

\bibitem{CSsimuls2}
M. Hindmarsh, S. Stuckey, and N. Bevis, ``Abelian Higgs Cosmic Strings: Small Scale Structure
and Loops'', Phys. Rev. D{\bf 79} (2009) 123504, [arXiv:0812.1929].\\
N.~Bevis, M.~Hindmarsh, M.~Kunz and J.~Urrestilla,
  ``CMB power spectrum contribution from cosmic strings using  field-evolution
  simulations of the Abelian Higgs model,''
  Phys.\ Rev.\  D {\bf 75}, 065015 (2007)
  [arXiv:astro-ph/0605018];

\bibitem{deficit}
A.~Vilenkin,
  ``Gravitational Field Of Vacuum Domain Walls And Strings,''
  Phys.\ Rev.\  D {\bf 23}, 852 (1981);\\
R.~Gregory,
  ``Gravitational Stability of Local Strings,''
  Phys.\ Rev.\ Lett.\  {\bf 59}, 740 (1987).

\bibitem{wake}
J.~Silk and A.~Vilenkin,
  ``Cosmic Strings And Galaxy Formation,''
  Phys.\ Rev.\ Lett.\  {\bf 53}, 1700 (1984);\\
 M. Rees,
 ``Baryon concentrations in string wakes at $z \geq 200$:
 implications for galaxy formation and large-scale structure,"
 Mon. Not. R. astr. Soc. {\bf{222}}, 27p (1986);\\
T.~Vachaspati,
  ``Cosmic Strings and the Large-Scale Structure of the Universe,''
  Phys.\ Rev.\ Lett.\  {\bf 57}, 1655 (1986);\\
A.~Stebbins, S.~Veeraraghavan, R.~H.~Brandenberger, J.~Silk and N.~Turok,
  ``Cosmic String Wakes,''
  Astrophys.\ J.\  {\bf 322}, 1 (1987);\\
E. Bertschinger,
 ``Cosmological Accretion Wakes,"
 Astrophys.\ J.\  {\bf 316}, 489 (1987).

\bibitem{wakeZel}
 L.~Perivolaropoulos, R.~H.~Brandenberger and A.~Stebbins,
  ``Dissipationless Clustering Of Neutrinos In Cosmic String Induced Wakes,''
  Phys.\ Rev.\  D {\bf 41}, 1764 (1990);\\
  R.~H.~Brandenberger, L.~Perivolaropoulos and A.~Stebbins,
  ``Cosmic Strings, Hot Dark Matter and the Large Scale Structure of the 
  Universe,''
  Int.\ J.\ Mod.\ Phys.\  A {\bf 5}, 1633 (1990).

\bibitem{zeldovich1}
Y.~.B.~Zeldovich,
  ``Gravitational instability: An Approximate theory for large density perturbations,''
  Astron.\ Astrophys.\  {\bf 5}, 84 (1970).

\bibitem{rhb1}
A.~Sornborger, R.~H.~Brandenberger, B.~Fryxell and K.~Olson,
  ``The Structure of cosmic string wakes,''
  Astrophys.\ J.\  {\bf 482}, 22 (1997)
  [astro-ph/9608020].
  
\bibitem{haramiyoshi}
T.~Hara and S.~Miyoshi,
  ``Formation Of The First Systems In The Wakes Of Moving Cosmic Strings,''
  Prog.\ Theor.\ Phys.\  {\bf 77}, 1152 (1987).

\bibitem{miyama1}
S.M. Miyama, S. Narita \& C. Hayashi,
\emph{Fragmentation of Isothermal Sheet-Like Clouds. I,}
Prog. Theor. Phys. {\bf 78} (1987), 1051.

\bibitem{fillmore1}
J.~A.~Fillmore and P.~Goldreich,
  ``Self-similiar gravitational collapse in an expanding universe,''
  Astrophys.\ J.\  {\bf 281}, 1 (1984).

\bibitem{reed1}
D.~Reed, R.~Bower, C.~Frenk, A.~Jenkins and T.~Theuns,
  ``The halo mass function from the dark ages through the present day,''
  Mon.\ Not.\ Roy.\ Astron.\ Soc.\  {\bf 374}, 2 (2007)
  [astro-ph/0607150].

\bibitem{bond1}
J.~R.~Bond, S.~Cole, G.~Efstathiou and N.~Kaiser,
  ``Excursion set mass functions for hierarchical Gaussian fluctuations,''
  Astrophys.\ J.\  {\bf 379}, 440 (1991).

\bibitem{Minkowski}
K.~R.~Mecke, T.~Buchert, H.~Wagner,
  ``Robust morphological measures for large scale structure in the universe,''
  Astron.\ Astrophys.\  {\bf 288}, 697-704 (1994).
  [astro-ph/9312028];\\
J.~Schmalzing and T.~Buchert,
  ``Beyond genus statistics: a unifying approach to the morphology of cosmic
  structure,''
  Astrophys.\ J.\  {\bf 482}, L1 (1997)
  [arXiv:astro-ph/9702130].

\bibitem{early2}
D.~Mitsouras, R.~H.~Brandenberger and P.~Hickson,
  ``Topological Statistics and the LMT Galaxy Redshift Survey,''
  arXiv:astro-ph/9806360;\\
H.~Trac, D.~Mitsouras, P.~Hickson and R.~H.~Brandenberger,
  ``Topology of the Las Campanas Redshift Survey,''
  Mon.\ Not.\ Roy.\ Astron.\ Soc.\  {\bf 330}, 531 (2002)
  [arXiv:astro-ph/0007125].

\bibitem{Evan}
E.~McDonough and R.~H.~Brandenberger,
  ``Searching for Signatures of Cosmic String Wakes in 21cm Redshift Surveys using Minkowski Functionals,''
  arXiv:1109.2627 [astro-ph.CO].

\bibitem{barkanaloeb1}
R.~Barkana and A.~Loeb,
  ``In the beginning: The First sources of light and the reionization of the Universe,''
  Phys.\ Rept.\  {\bf 349}, 125 (2001)
  [astro-ph/0010468].

\bibitem{pat1}
P.~Scott and S.~Sivertsson,
  ``Gamma-Rays from Ultracompact Primordial Dark Matter Minihalos,''
  Phys.\ Rev.\ Lett.\  {\bf 103}, 211301 (2009)
  [Erratum-ibid.\  {\bf 105}, 119902 (2010)]
  [arXiv:0908.4082 [astro-ph.CO]].

\bibitem{Wangyi}
  O.~F.~Hernandez, Y.~Wang, R.~Brandenberger and J.~Fong,
  ``Angular 21 cm Power Spectrum of a Scaling Distribution of Cosmic String
  Wakes,''
 JCAP {\bf 1108}, 014 (2011)
  [arXiv:1104.3337 [astro-ph.CO]].

\bibitem{Oscar}
O.~F.~Hernandez and R.~H.~Brandenberger,
  ``The 21 cm Signature of Shock Heated and Diffuse Cosmic String Wakes,''
  JCAP {\bf 1207}, 032 (2012)
  [arXiv:1203.2307 [astro-ph.CO]].

\bibitem{Pagano}
M.~Pagano and R.~Brandenberger,
  ``The 21cm Signature of a Cosmic String Loop,''
  JCAP {\bf 1205}, 014 (2012)
  [arXiv:1201.5695 [astro-ph.CO]].

\end{thebibliography}
\end{document}